# Assessing the drag implications of hydrogen fuel utilization in small aircraft: a preliminary numerical analysis

Serena Zhang[1], Dong Eun Lee[2], Li Qiao[2]

[1]*The Bishop's School, La Jolla. CA, 92037*
[2]*School of Aeronautics and Astronautics, Purdue University, West Lafayette, IN 47907*

## Abstract

The urgent need for cleaner, sustainable aviation fuels has driven interest in the use of hydrogen as a potential replacement for traditional hydrocarbon fuels in aviation. However, transitioning to hydrogen presents significant challenges due to its lower volumetric energy density which requires a larger fuel volume for equivalent energy. This study uses Computational Fluid Dynamics (CFD) simulations, utilizing the ANSYS Fluent 2022 R1 solver, to conduct a preliminary investigation on the aerodynamic implications of adopting hydrogen fuel in small aircraft, specifically the Bonanza G36. Three different types of hydrogen fuel - gaseous hydrogen at 350 bar, gaseous hydrogen at 700 bar, and liquid hydrogen - were considered. Despite a promising 64% reduction in fuel weight in comparison to gasoline, the volume required for hydrogen fuel was found to increase significantly, with values ranging from 263% to 896% compared to the base case. The aerodynamic impact was quantified in terms of drag, with the study revealing increases of 67.32%, 48.42%, and 27.05% for gaseous hydrogen at 350 bar, gaseous hydrogen at 700 bar, and liquid hydrogen respectively. While the findings present a challenge, they also set a platform for exploring innovative fuel tank designs and drag reduction technologies to mitigate these effects. This study underscores the complexities involved in transitioning to hydrogen fuel in aviation, contributing valuable insights to guide further research towards sustainable aviation.

## Introduction

In recent years, the ever-increasing climate crisis has necessitated sustainable solutions across all industries. Particularly, the booming aviation industry has sought out alternative fuels to replace conventional jet fuels. In fact, transportation and energy are the two largest contributors to carbon emissions. Aviation constitutes 12% of all carbon emissions from transportation and 3% of the overall contributing factors to global warming; although this may seem insignificant, by 2050, it is predicted that aviation will contribute beyond 10% of overall emissions, a number that cannot be ignored [1]. While current answers involve setting industry requirements for CO2 emissions on major commercial aircraft for prominent airlines, researchers are exploring alternative fuels to find more innovative, proactive solutions to the energy crisis. Hydrogen, known for its high specific energy and non-carbon-emissive combustion, has emerged from this discourse as a promising candidate. Not only has hydrogen headlined other sectors of sustainable energy, such as nuclear fusion energy generators utilizing hydrogen isotopes [2], but hydrogen could potentially power commercial aircraft, reducing carbon emissions drastically. However, the practical implementation of hydrogen fuels poses challenges of its own: first, its lower volumetric energy density compared to traditional petroleum fuels [3]. To achieve an equivalent energy to that of its jet-fuel counterparts, hydrogen fuel requires tanks of more volume. Consequently, this increase in the necessary fuel volume could potentially interfere with the aerodynamics of the aircraft, threatening the stability and reliability of the plane. Most notably, altering the shape of the aircraft could increase the drag generated, which would subsequently increase the energy needed to overcome this extra drag force. This places excess stress on the aircraft's energy abilities.

Recent studies, such as those conducted by Verstraete et al. [4] and Zhao et al. [5], highlight the potential of hydrogen as a sustainable fuel alternative due to its high specific energy and zero-carbon emission combustion. These qualities make it a compelling candidate for future aircraft propulsion. However, the challenge of hydrogen's lower volumetric energy density, requiring a larger fuel tank compared to conventional fuels, has also been recognized [6, 7].

Other research has examined the impact of larger fuel volumes on aircraft performance. Notably, a body of work has emerged focusing on the effects of large fuel volumes on drag force. Sargeant et al. [8] and Brown et al. [9] have conducted detailed examinations of the role of drag in aviation fuel efficiency, suggesting that the alteration of drag due to larger fuel tanks could pose a substantial challenge in the transition to hydrogen fuel.

However, existing literature primarily investigates large commercial and freight aircraft [9, 10], leaving a knowledge gap in understanding the specific implications of hydrogen fuel for smaller aircraft. This brings us to the objective of our study: to investigate, numerically, the impact of an enlarged hydrogen fuel tank on the drag forces experienced by a small aircraft — specifically, the Bonanza G36 model [11].

The goal of this investigation is to deepen the understanding of the implications of hydrogen fuel use in small aircraft, providing actionable insights to engineers and policymakers alike. By mapping the impacts of a larger fuel tank on aircraft aerodynamics and performance, this study aims to contribute valuable knowledge to the ongoing efforts to transition the aviation industry towards sustainable fuel alternatives, thereby paving the way for a more sustainable aviation future.

## Methods

### Base model selection

The selection of an appropriate base model is critical for accurately evaluating the impacts of hydrogen fuel on the aerodynamic performance of small aircraft. In this study, the Bonanza G36, a small general aviation aircraft, was selected as the base model due to its size, popular usage, and well-documented specifications.

The Bonanza G36, manufactured by the Beechcraft Corporation, is a single-engine, piston aircraft widely used in both personal and commercial aviation. Its popularity in the aviation industry ensures that the study findings will be broadly applicable and have a significant impact. Figure 1 provides a visual representation of the Bonanza G36.

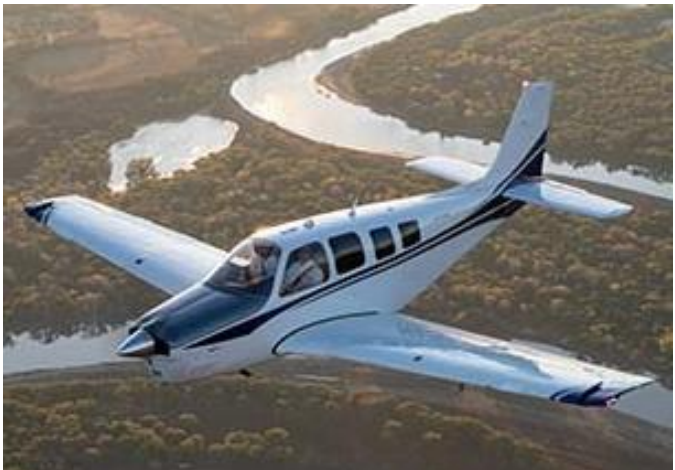

*Figure 1. A picture of Bonanaza G36.*

*Table 1. Bonanaza G36 specifications.*

| Parameter | Value |
|---|---|
| Length (m) | 8.38 m |
| Height (m) | 2.62 m |
| Wingspan (m) | 10.21 m |
| Wing Area (m$^2$) | 16.82 m$^2$ |
| Maximum Occupants | 6 |
| Maximum Cruise Speed (km/h) | 322 (≈ 90 m/s) |
| Maximum Range (km) | 1704 km |

In terms of specifications, the Bonanza G36 features a length of 8.38 meters, a wingspan of 10.21 meters, and a maximum takeoff weight of 1650 kilograms [12]. These specifications make it representative of a range of small, general aviation aircraft, enhancing the relevance of the study's findings. Detailed specifications of the Bonanza G36 can be found in Table 1.

To accurately calculate the drag force changes due to the enlarged volume of the fuel tank, a CAD model was created, replicating the Bonanza G36 in high fidelity. Figure 2 shows this CAD model. The use of a CAD model allowed for a detailed analysis of the potential changes in the aerodynamic performance and allowed for a more precise numerical investigation.

By focusing on a commonly used and representative aircraft model, this study provides meaningful insights into the practical implications of transitioning small aircrafts to hydrogen fuel. The selection of the Bonanza G36 enhances the applicability of the research findings and paves the way for further detailed studies in the realm of sustainable aviation.

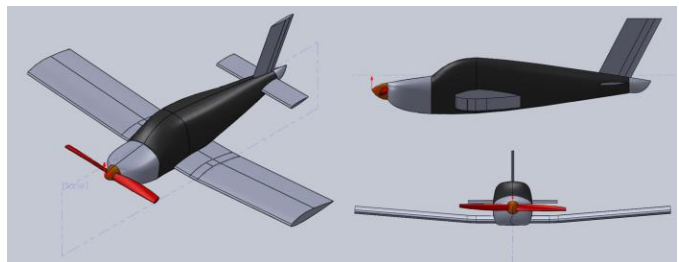

*Figure 2. 3D view of the base model (Bonanaza G36 replica).*

### *Hydrogen fuel consideration*

When contemplating the transition to hydrogen as a fuel for aviation, the unique properties of hydrogen must be evaluated. The gravimetric and volumetric energy densities of hydrogen differ significantly from conventional gasoline, impacting both the weight and volume of fuel necessary for equivalent energy output.

Hydrogen is renowned for its high gravimetric density, meaning it delivers a considerable amount of energy per unit of mass. This property is visually illustrated in Figure 3, which depicts the gravimetric and volumetric energy densities of various common fuels. When comparing hydrogen to gasoline, it is evident that hydrogen's gravimetric energy density far surpasses that of gasoline.

This superior gravimetric density of hydrogen translates into a significant reduction in fuel weight. For instance, while the base model Bonanza G36 fueled by gasoline requires 201 kg of fuel, the equivalent energy output can be achieved with only 72 kg of hydrogen. This amounts to a substantial 64% decrease in fuel weight, which in turn leads to lower aircraft weight. The benefits of such a reduction cannot be overstated, as a lighter aircraft inherently requires less energy to achieve and maintain flight, potentially leading to further improvements in energy efficiency and operational cost savings.

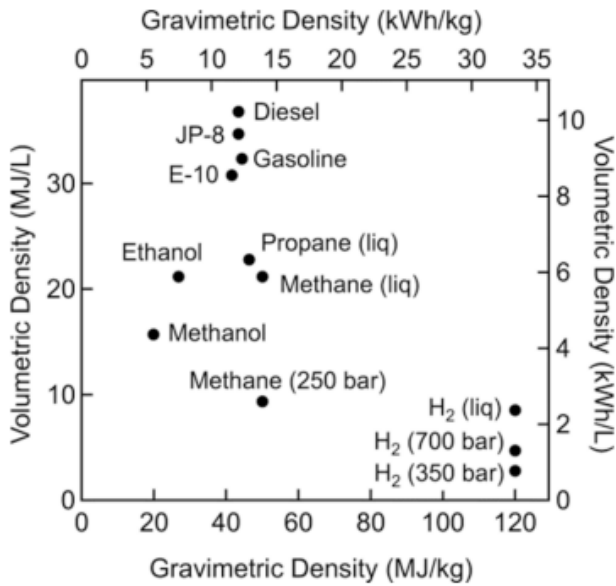

*Figure 3. Gravimetric and volumetric energy densities of common fuels based on lower heating values calculated for ambient temperature conditions (Reference [1]).*

Despite these advantages, the use of hydrogen fuel also presents some challenges, primarily due to its lower volumetric density compared to gasoline. Our calculations revealed the need for a significantly larger fuel tank to accommodate the gasoline energy equivalent in hydrogen fuel. Three different types of hydrogen fuel were considered in this study: gaseous hydrogen at 350 bar, gaseous hydrogen at 700 bar, and liquid hydrogen. Each type presents a different volumetric density, which affects the required fuel tank size. The specific results can be found in Table 2.

For gaseous hydrogen at 350 bar, an additional fuel volume of 2508 L is needed, marking a remarkable 896% increase. Gaseous hydrogen at 700 bar requires 1449 L additional fuel, signifying a 518% increase. Even when considering liquid hydrogen, with its higher volumetric density, an additional 737 L of fuel volume is required. This figure still corresponds to a substantial 263% increase compared to the base model.

From these findings, it becomes clear that while the transition to hydrogen fuel presents promising advancements in sustainable aviation, such as considerable weight reduction in fuel, it also introduces the challenge of accommodating a larger fuel tank. This reality necessitates a careful consideration of the trade-offs involved in implementing hydrogen fuel. These findings also highlight the need for innovative solutions to integrate larger fuel storage without compromising the aircraft's aerodynamic performance.

## Hydrogen tank configuration

The configuration of the fuel tank is a pivotal factor in considering the practical transition from conventional fuels to hydrogen, given the latter's specific properties such as lower volumetric energy density. There are typically three main types of fuel tank configurations to be considered: external, internal, and hybrid.

External fuel tanks are often used to increase the fuel capacity of an aircraft without modifying the internal structure. However, their external placement can influence an aircraft's aerodynamics, potentially increasing drag and altering flight characteristics [13]. On the positive side, external tanks are relatively easy to install and remove, especially in case of repairs, providing a degree of flexibility.

Internal fuel tanks, on the other hand, are integrated within the aircraft's structure. These tanks offer the advantage of maintaining the aircraft's exterior shape, hence not altering its aerodynamic profile [14]. However, internal tanks require substantial structural modifications and can limit available space for other aircraft systems or cargo.

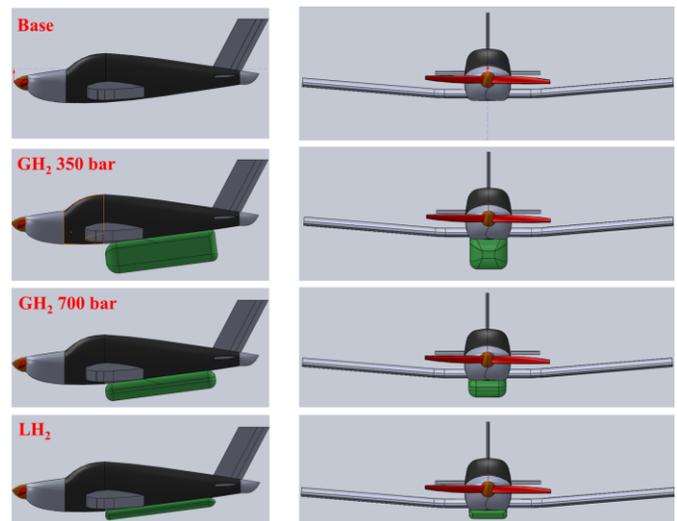

*Figure 4. Fuel tank configuration for $GH_2$ and $LH_2$.*

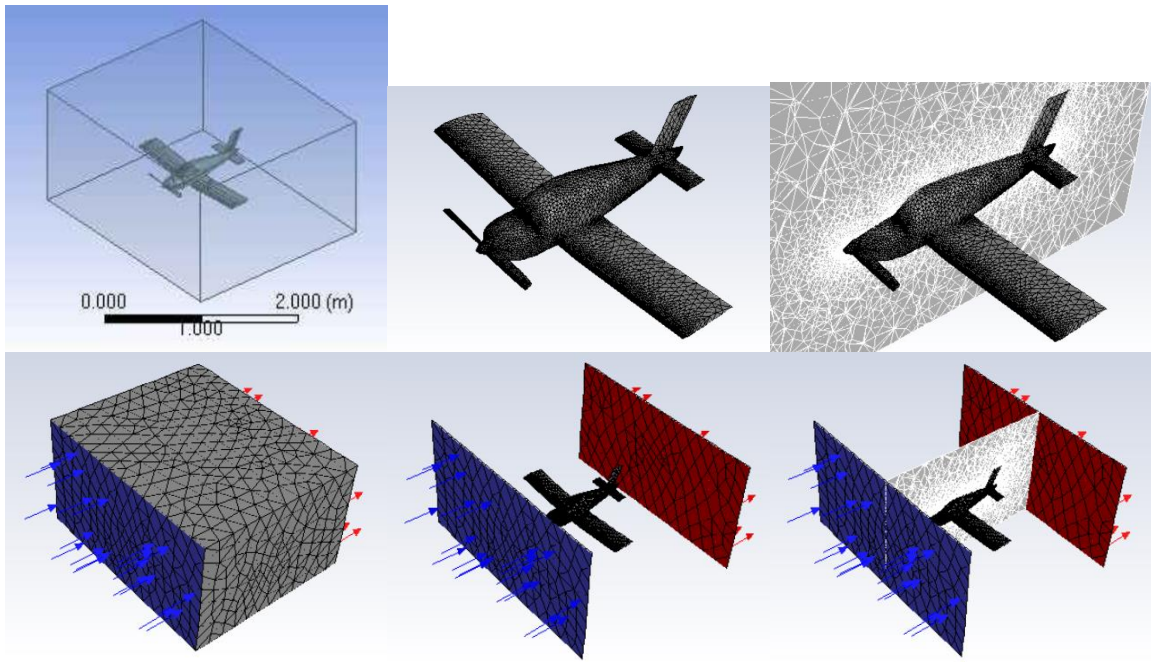

*Figure 5. CFD domain, geometry, and mesh used in the study.*

Hybrid configurations combine the advantages of both external and internal tank designs. They allow for increased fuel capacity and can maintain a favorable aerodynamic profile, but they also require more complex design and integration processes [15].

For the purpose of this study, we adopted an external fuel tank configuration, specifically one situated under the fuselage. This decision was driven by simplicity in design modification and iteration. While this location may not be optimal as it could interfere with the landing gears, it is a suitable choice for a preliminary study investigating the impact of hydrogen fuel on generated drag. Detailed representations of different external fuel tank configurations for each hydrogen fuel type are compared to the base model in Figure 4.

It is worth mentioning that different types of hydrogen fuel ($GH_2$ at 350 bar, $GH_2$ at 700 bar, and $LH_2$) necessitate different storage requirements and sub-systems for fuel delivery. For instance, liquid hydrogen demands robust insulation to maintain its extremely low temperature, while high-pressure gaseous hydrogen requires strong tank walls to withstand the pressure [16]. Nevertheless, in this study, we focused on the pure volume of fuel without considering tank thickness, insulation or the specifics of the fuel delivery system. While a more comprehensive study may delve into these details, this simplification allows us to focus on the fundamental question of drag implications.

In summary, while the external fuel tank configuration employed in this study presents certain challenges and limitations, it serves as a valuable starting point to examine the impact of a larger fuel tank on the aerodynamic performance of small aircraft. Future work could refine this configuration and explore other potential tank configurations, considering the specific storage requirements for different forms of hydrogen fuel.

### *Geometry and mesh*

The creation of accurate geometries and meshes forms the foundation for reliable computational fluid dynamics (CFD) simulations. For this study, the aircraft model was scaled to a 1:6.5 ratio of the actual Bonanza G36 size, primarily to reduce the geometric size and simulation time. The scaled model was then placed inside a 1 x 1 x 1 m enclosure, which served as a controlled environment for setting the boundary conditions.

The geometries and meshes were created using the ANSYS Fluent 2022 R1 Workbench software. The mesh consisted of tetrahedral cells, and depending on the specific fuel type under consideration, the cell count ranged from approximately 1,270,000 to 1,400,000. This cell size ensured a suitable compromise between computational efficiency and resolution, allowing for accurate capture of flow details while maintaining manageable computation times. Figure 5 offers a visual representation of the CFD domain, geometry, and mesh utilized in the study.

### *Boundary conditions*

The boundary conditions are a critical factor in obtaining valid and meaningful simulation results. In this study, we opted for a velocity inlet and pressure outlet condition. The inlet velocity was set to match the maximum cruising speed of the Bonanza G36, which is approximately 90 m/s. At the outlet, a pressure condition of 0 Pa Gauge was employed, representing an open-air environment.

Both the enclosure wall and the airplane surfaces were set to a no-slip condition, which is a standard approach in fluid dynamics simulations to reflect the zero velocity at a solid boundary. The working fluid was air, given the application context of aviation. Figure 6 visualizes how the enclosure and airplane were arranged under the specified boundary conditions, with the blue plane on the left-hand side representing the inlet, and the red plane on the right-hand side depicting the outlet.

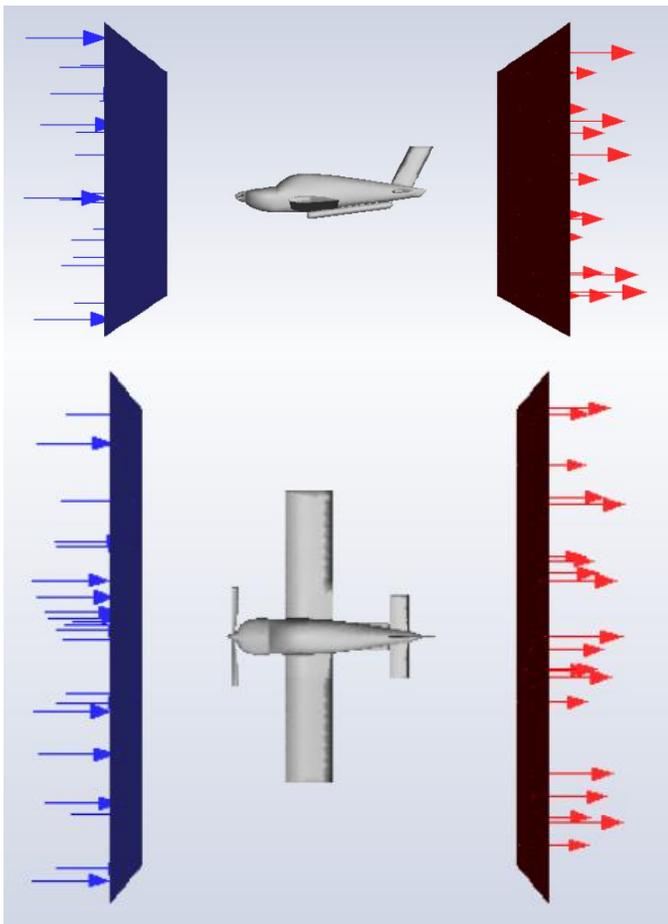

*Figure 6. Boundary conditions (blue: velocity inlet, red: pressure outlet)*

### *Numerical setup*

The numerical setup for this study was configured using the ANSYS Fluent 2022 R1 solver. We opted for a 3D setup, a pressure-based solver, and a steady solver. These choices are specifically tailored to the nature of our investigation.

The 3D setup allows for a realistic and detailed representation of the flow phenomena around the aircraft. While 2D simulations can provide useful insights and are computationally less expensive, they are inherently limited in their ability to capture the complex, three-dimensional nature of real-world fluid flows.

A pressure-based solver was employed as it is well-suited for low-speed and incompressible flows, typical for aircraft in cruising conditions.

The choice of a steady solver is a common approximation in aircraft performance analysis. While some flow phenomena are unsteady, an aircraft cruising at a constant altitude and velocity reaches a relatively steady state, making this an appropriate simplification.

For turbulence modeling, we employed the k-epsilon Realizable model. This two-equation model has been widely used in aerodynamics due to its balance of computational efficiency and predictive capability. It includes the effects of streamline curvature and swirl, which are particularly relevant to the complex flow around an aircraft and its fuel tanks.

The solver used the Coupled Scheme for pressure-velocity coupling, which provides increased stability and convergence speed for complex flows. Spatial discretization was conducted using the Second Order Pressure and Second Order Upwind schemes for Momentum, Turbulent Kinetic Energy, and Turbulent Dissipation Rate. The second-order schemes are more accurate than first-order ones, reducing numerical diffusion and better preserving the flow details.

The described numerical setup allows for accurate and reliable CFD simulations of the aerodynamic performance of the aircraft models under the influence of various hydrogen fuel tank configurations. Through this, it provides the necessary basis for the investigation of drag force alterations due to the transition to hydrogen fuel.

### **Results**

### *Convergence*

In this study, the convergence was assessed using scaled residuals for the x, y, and z-velocity components, as presented in Figure 7. As can be observed, the residuals dropped below the chosen convergence criterion of 1e-06 after approximately 200 iterations. This indicates that the solution reached a consistent state where further iterations would not significantly alter the results.

Figure 8 provides additional insight into the convergence by monitoring the drag and drag coefficient values throughout the iterative process. As these values stabilized after approximately 200 iterations, it provides further assurance of the solution's convergence.

Achieving a good level of convergence is important as it not only guarantees the accuracy of the results, but it also ensures the stability of the solution. A well-converged solution means that the results reflect a true steady state and are not influenced by initial conditions or numerical artifacts.

Therefore, based on these findings, we can state with confidence that the simulations have reached a satisfactory level of convergence, thus providing a reliable foundation for subsequent analysis and discussion of the results.

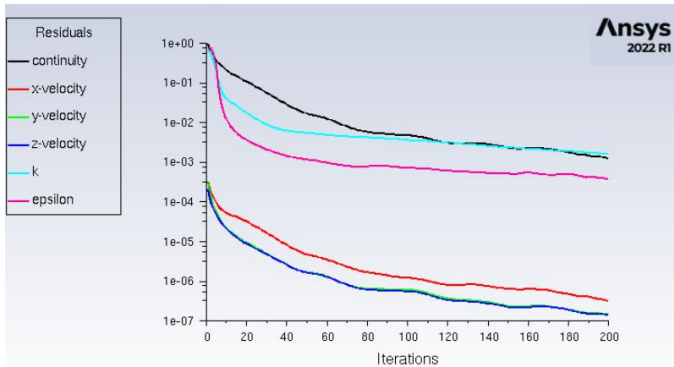

*Figure 7. Plot of scaled residuals.*

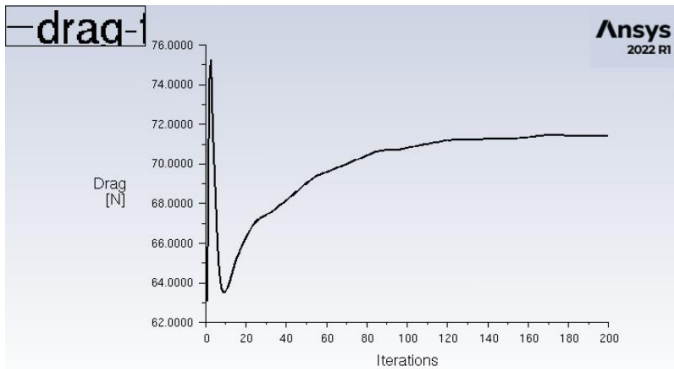

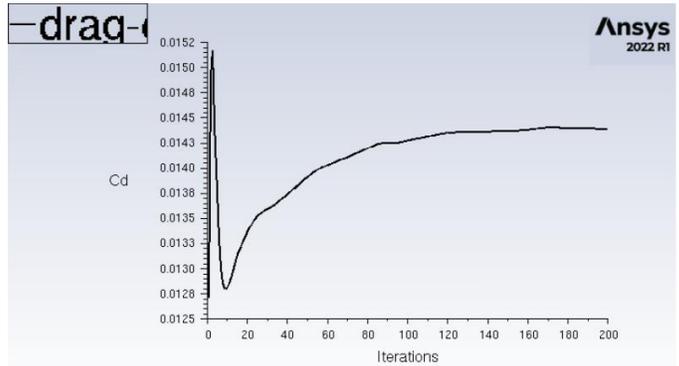

*Figure 8. Plots of Drag and Drag Coefficient values throughout the iterative process.*

### *Effect of hydrogen fuel tank on drag force*

In our computational analysis, the impact of hydrogen fuel tanks on the aerodynamic performance of the aircraft, particularly on the drag force, was assessed. Table 3 provides the numerical results for the three different hydrogen fuel cases compared with the base case, which used gasoline as fuel.

The implementation of an external hydrogen fuel tank under the fuselage has proven to increase the drag force produced by each case of hydrogen fuel. A prominent rise in drag force was seen in the $GH_2$ 350 bar case, where the force surged to 119.51 N, marking a substantial increase of 67.32% compared to the base case drag of 71.43 N. This is a significant rise in the aerodynamic drag which can adversely impact the aircraft's performance, notably in its fuel efficiency and range, due to the increased energy required to overcome the heightened drag force.

For the $GH_2$ 700 bar case, the drag force reduced to 106.01 N, signifying an increase of 48.72% compared to the base case. However, this is a lesser increase than seen in the $GH_2$ 350 bar case. The reduction in drag force compared to the 350 bar case can be attributed to the higher energy density of hydrogen at 700 bar, leading to a comparatively smaller fuel tank and thereby lesser disruption of the aircraft's aerodynamics.

*Table 3. CFD results for $GH_2$ and $LH_2$ considerations.*

|  | Gasoline 100LL (Base) | $GH_2$ 350 bar | $GH_2$ 700 bar | $LH_2$ |
|---|---|---|---|---|
| **Drag (N)** | 71.43 | 119.51 | 106.01 | 90.75 |
| **Drag coefficient** | 0.014 | 0.024 | 0.021 | 0.018 |
| **Lift (N)** | 279.72 | 388.14 | 326.05 | 270.21 |
| **Lift coefficient** | 0.056 | 0.078 | 0.066 | 0.054 |
| **Lift-to-Drag ratio** | 3.92 | 3.25 | 3.08 | 2.98 |
| **Change of drag force from the base (%)** | 0.00 | 67.32 | 48.42 | 27.05 |

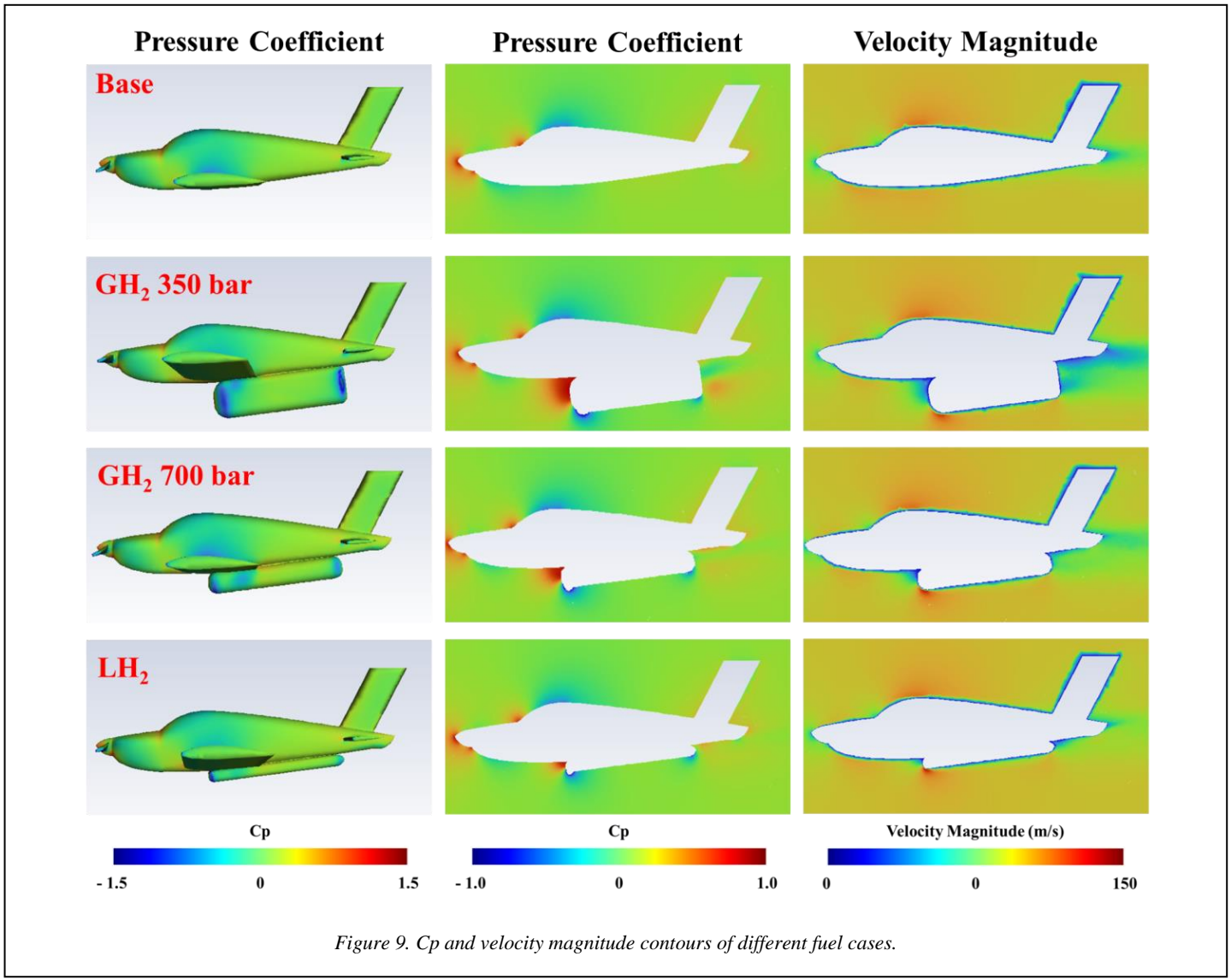

*Figure 9. Cp and velocity magnitude contours of different fuel cases.*

When considering the LH$_2$ case, the drag force further reduced to 90.75 N, translating to a 27.05% increase in drag force compared to the base case. While this is still a noticeable rise in drag, it becomes more justifiable when considering the potential benefits of hydrogen fuel. Hydrogen fuel boasts the benefits of clean combustion, emitting only water vapor, which has massive implications for environmental sustainability. Unlike conventional fuels, hydrogen fuels will not emit carbon dioxide and contribute to the overall carbon emissions of the aviation industry. Moreover, hydrogen's superior gravimetric energy density results in a much lighter fuel load than gasoline fuels, potentially offsetting some of the performance drawbacks associated with increased drag.

Thus, although the use of hydrogen fuel in its different forms invariably led to an increase in drag force, this rise can be mitigated through optimization strategies such as efficient fuel tank design and positioning. In the context of sustainable aviation and reduced carbon emissions, the benefits of using hydrogen fuel seem promising enough to warrant further exploration and research.

To further understand the reasons for the observed differences in drag, we explored the Pressure Coefficient (Cp) and Velocity Magnitude contours as presented in Figure 9. This figure presents these contours for four different fuel cases: base, GH$_2$ 350 bar, GH$_2$ 700 bar, and LH$_2$.

As indicated in the figure, the introduction of the external fuel tank under the fuselage results in high Cp regions right upstream of the fuel tank head. These high-pressure regions are a direct consequence of the interruption of the flow around the fuselage due to the externally mounted fuel tank. As a result, a pressure buildup occurs, contributing to an increase in drag.

Interestingly, these high Cp regions shrink in size with the increase in pressure and the transition to liquid hydrogen. This can be attributed to the lesser volume required for storing higher pressure gaseous hydrogen and liquid hydrogen, leading to smaller fuel tanks and subsequently smaller high-pressure zones.

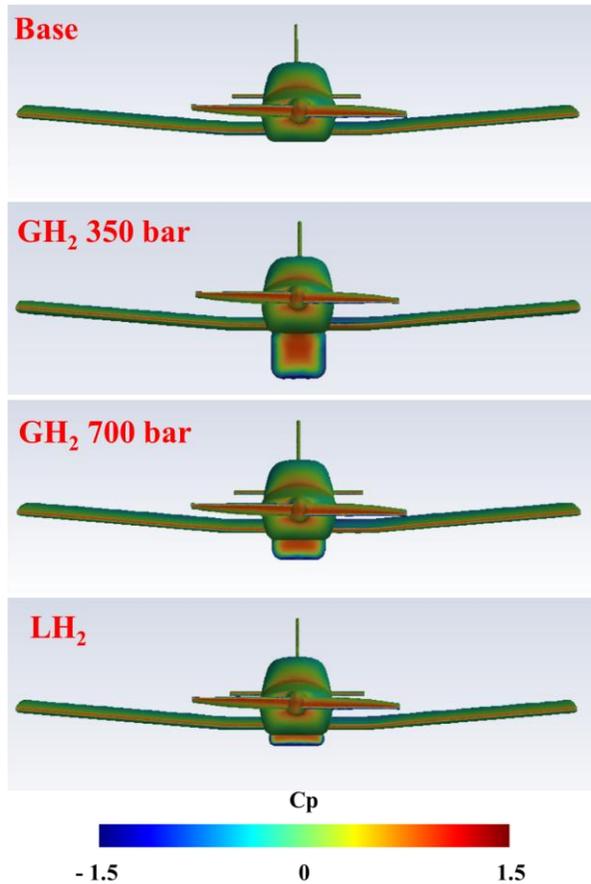

*Figure 10. Frontal-view Cp contour of different fuel cases.*

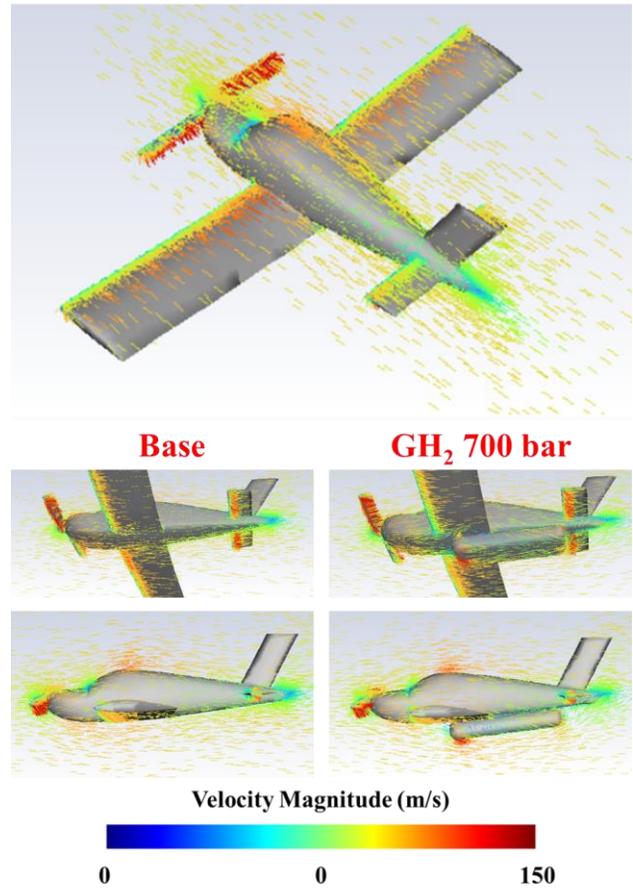

*Figure 11. Velocity vectors around the aircraft of different fuel cases.*

Figure 10, which shows the frontal view of Cp contours on the aircraft's surface, provides further clarity. It visibly shows that the high Cp regions are concentrated on the frontal face of the fuel tank and decrease in size as we move from lower pressure gaseous hydrogen to higher pressure and then to liquid hydrogen. This observation is consistent with our previous findings, where higher drag forces are associated with larger high-pressure zones.

These visualizations not only help to explain the increase in drag observed with the addition of the external fuel tank but also suggest that more efficient tank design and placement could help to mitigate these effects.

The examination of the velocity vectors around the airplane, as demonstrated in Figure 11, further elucidates the flow characteristics influenced by the presence of the external hydrogen fuel tank. From the top-view, the flow around the aircraft seems largely unaffected, which can be attributed to the fact that no major design changes have been implemented at the top of the aircraft.

However, the bottom- and side-views paint a starkly different picture. The flow lines around the aircraft clearly show that the introduction of the hydrogen fuel tank under the fuselage disturbs the streamline flow. This disruption is expected to induce additional drag due to flow separation and increase in turbulent kinetic energy around the tank. The flow visualization further reinforces the previous observation from the Cp contours, demonstrating the direct impact of the external fuel tank on the overall drag force.

This flow interruption around the aircraft due to the external tank's presence necessitates careful consideration and design refinement. Innovative approaches towards fuel tank design and placement, such as conformal fuel tanks or distributed fuel storage within the structure, could potentially lead to a reduction in the observed drag increases. Furthermore, application of flow control techniques or drag reduction technologies could be instrumental in overcoming the aerodynamic challenges posed by the adoption of hydrogen fuel.

In summary, while the switch to hydrogen fuel in aviation presents promising opportunities for sustainability and emission reduction, it also brings with it certain challenges. The increase in drag due to the larger fuel tank requirements for hydrogen storage is a significant one, but with thorough understanding, appropriate design interventions, and further research, it can be addressed. This study provides a crucial step in that direction, deepening our understanding of the aerodynamic implications of hydrogen fuel use in small aircraft, and laying a foundation for more sustainable aviation solutions.

## Discussion

This study shed light on the complex challenges of adopting hydrogen as a potential aviation fuel, focusing on small aircraft design. The primary challenge that hydrogen fuels pose is that the lower volumetric energy density of hydrogen requires a larger fuel volume for an equivalent energy output compared to traditional gasoline fuels, which has significant aerodynamic implications. Despite achieving a substantial 64% reduction in fuel weight by adopting hydrogen fuel, the study revealed an unavoidable increase in fuel volume, which ranged from 263% to 896% depending on the type of hydrogen fuel used.

The aerodynamic implications of these increased volumes were evaluated through CFD simulations. These simulations revealed increased drag forces on the aircraft due to the presence of an external fuel tank, which altered the external geometry of the aircraft. The study quantified these increases to be 67.32%, 48.42%, and 27.05% for gaseous hydrogen at 350 bar, gaseous hydrogen at 700 bar, and liquid hydrogen, respectively.

While these findings present an apparent challenge in the transition to sustainable aviation fuels, they simultaneously highlight opportunities for future research and development. Specifically, the need for innovative fuel tank designs and drag reduction technologies is imperative to counteract these aerodynamic disadvantages.

The implications of this study were limited by the variations of aircraft and fuel tank designs that were tested. While these results have direct implications for small aircraft such as the Bonanza G36 used in the CFD simulations, larger commercial aircraft are greater contributors to carbon emissions and should be the next target of this research. This study utilized an external tank configuration, but if other tank configurations, such as internal and hybrid, were considered for commercial aircrafts, the external shape of the aircraft may not be as dramatically altered, but other elements of the aircraft may be affected. For instance, the main cabin room used for seating may need to be set aside for more fuel storage space. These additional cases should be examined in order to Overall, the findings of this study are still significant, as the implementation of hydrogen fuel must begin on a small scale; once hydrogen is adopted successfully by small aircraft, it can be more widely utilized in larger commercial aircraft.

Ultimately, this study provides valuable insights into the complexities of transitioning to hydrogen fuel and underscores the importance of multi-disciplinary solutions to achieve a sustainable future in aviation. While there are certain structural and aerodynamic barriers to immediate adoption of hydrogen fuels, the findings from this study point towards important areas of research, which once explored, can lead to promising strategies to transition the aviation industry towards sustainable fuels and climate-friendly solutions.